\documentclass{iau379}

\usepackage{amsmath}
\usepackage{graphicx}
\usepackage{multirow}

\usepackage{fontawesome5}
\usepackage{hyperref}
\usepackage{xcolor}
\usepackage{amsmath}
\usepackage{gensymb}
\usepackage{enumitem}
\usepackage{amsfonts}

\graphicspath{{./}{figures/}}
\hypersetup{colorlinks,urlcolor=blue}

\newcommand{\github}[1]{%
   \href{#1}{\textcolor{black}\faGithubSquare}%
}

\def\PhySO{{$\Phi$-SO}}

\def\placeholder{{\square}}
\def\LCDM{{$\Lambda$CDM}}

\def\ltsima{$\; \buildrel < \over \sim \;$}
\def\simlt{\lower.5ex\hbox{\ltsima}}
\def\gtsima{$\; \buildrel > \over \sim \;$}
\def\simgt{\lower.5ex\hbox{\gtsima}}

\def\dotsfill{\leaders\hbox to 1em{\hss.\hss}\hfill}

\def\Myr{{\rm\,Myr}}

\def\s{\ifmmode \widetilde \else \~\fi}
\def\={\overline}

\def\spose#1{\hbox to 0pt{#1\hss}}

\def\eg{{e.g.,\ }}
\def\ie{{i.e.\ }}

\begin{document}

\lefttitle{Tenachi et al.}
\righttitle{Recovering a free-form
potential from stellar coordinates}

\jnlPage{1}{7}
\jnlDoiYr{2023}
\doival{10.1017/xxxxx}

\aopheadtitle{Proceedings of IAU Symposium 379}
\editors{P. Bonifacio,  M.-R. Cioni \& F. Hammer, eds.}

\title{An end-to-end strategy for recovering a free-form potential from a snapshot of stellar coordinates}

\author{Tenachi W. $^1$, Ibata R. $^1$, Diakogiannis F. $^2$}
\affiliation{$^1$ Universit\'e de Strasbourg, CNRS, Observatoire astronomique de Strasbourg, UMR 7550, F-67000 Strasbourg, France\\
$^2$ Data61, CSIRO, Kensington, WA 6155, Australia}

\begin{abstract}
New large observational surveys such as Gaia are leading us into an era of data abundance, offering unprecedented opportunities to discover new physical laws through the power of machine learning. Here we present an end-to-end strategy for recovering a free-form analytical potential from a mere snapshot of stellar positions and velocities. First we show how auto-differentiation can be used to capture an agnostic map of the gravitational potential and its underlying dark matter distribution in the form of a neural network. However, in the context of physics, neural networks are both a plague and a blessing as they are extremely flexible for modeling physical systems but largely consist in non-interpretable black boxes. Therefore, in addition, we show how a complementary symbolic regression approach can be used to open up this neural network into a physically meaningful expression. We demonstrate our strategy by recovering the potential of a toy isochrone system.
\end{abstract}


\maketitle

\section{Introduction}

The Lambda cold dark matter (\LCDM) model is very successful at reproducing large scale observations. However, it poses several challenges at the galactic scale \citep{Bullock_LCDM_challenges}. Being able to compute a reliable and high resolution map of the dark matter distribution from observations in our galactic neighborhood would be of great value for enabling us to decide on these issues.
This challenging task is being rendered feasible by the European Space Agency’s Gaia mission which is measuring the distance and radial velocity of tens of millions of stars, enabling us to have access for the first time to a very large dataset of 6D (position and velocity) stellar coordinates \citep{GaiaDR3}.

Here we present a novel strategy for mapping the galactic potential and encapsulating it in an analytical expression in an agnostic and model free manner. In Section \ref{sec:MassFinder} we give an auto-differentiation based framework for recovering a free form neural network model of the potential from the snapshot of frozen stellar coordinates that we are able to measure at present time. Then in Section \ref{sec:SymbolicRegression}, we show how our reinforcement learning based symbolic regression framework \PhySO\ \citep{physo} can be used to distill the captured neural network potential model into an \emph{a priori} unknown but physically meaningful functional form by searching in the space of equations.

\section{Learning a potential from a snapshot positions and velocities}
\label{sec:MassFinder}

\begin{figure*}[h]
\begin{center}
\includegraphics[angle=0, clip, width=\hsize]{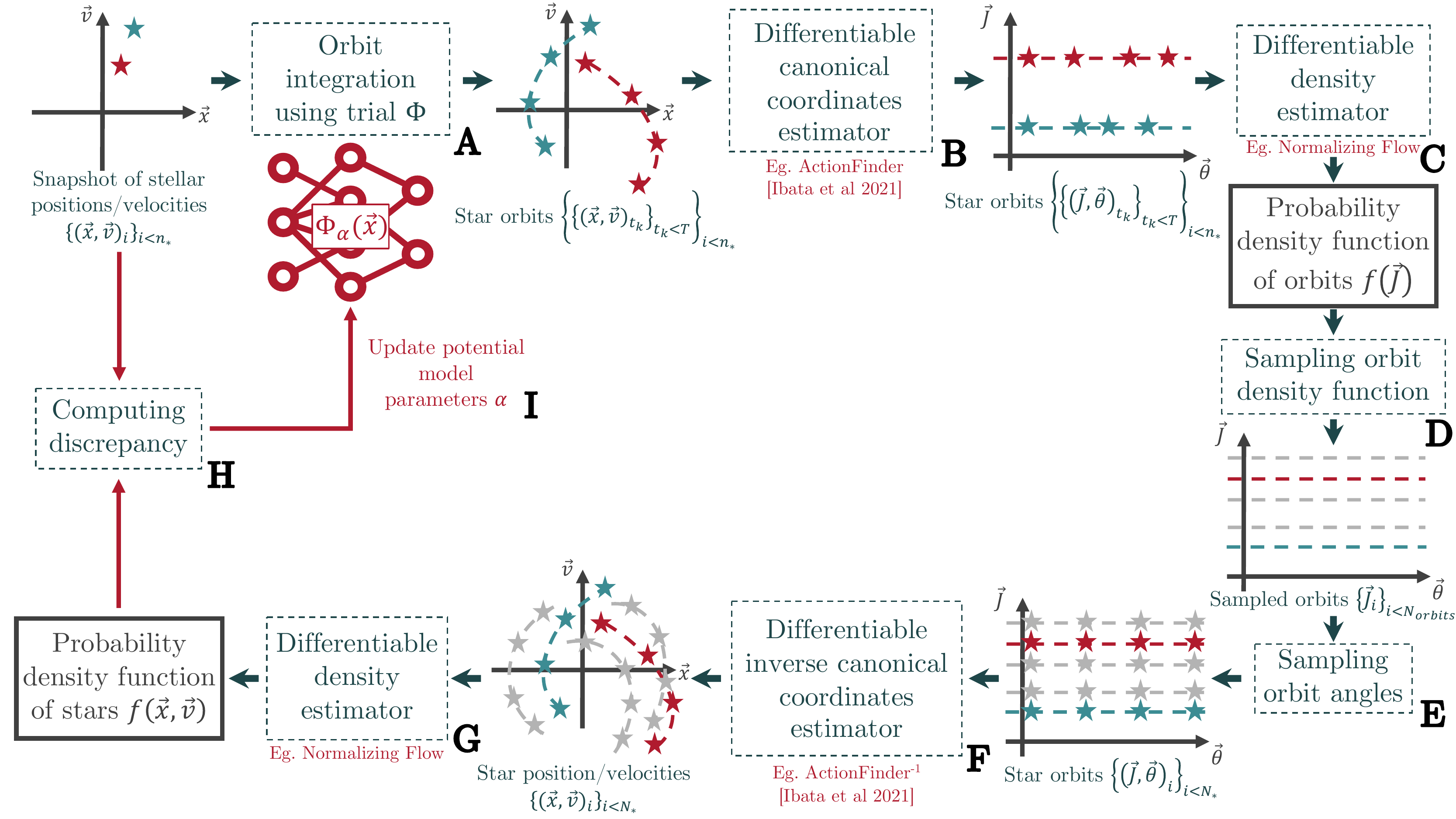}
\end{center}
\caption{Proposed strategy for recovering a free form neural network potential that stabilizes a distribution of stars $\{(\vec{x}, \vec{v})_i\}_{i<n_*}$ using auto-differentiation and gradient descent. See \S 1 of Section \ref{sec:MassFinder} for a full description of the workflow.}
\label{fig:MassFinder}
\end{figure*}

The framework we propose for agnostically recovering the potential is shown in Figure~\ref{fig:MassFinder}. The input data of this framework consists of phase-space stellar coordinates $\{(\vec{x}, \vec{v})_i\}_{i<n_*}$ obtained from catalogues such as Gaia. 

\begin{enumerate}[label=\Alph*.]

\item These stars are integrated in a trial gravitational potential $\Phi_\alpha$ represented by a flexible free form neural network that depends on parameters $\alpha$. 

\item These trajectories can then used to deduce orbits in action space using a differentiable canonical coordinates estimator. For this purpose, the neural network based \texttt{ACTIONFINDER} method (\cite{ActionFinder}) can be used. This transformation to the space of orbits represented by three integrals of motions \ie actions : $\Vec{J} = (J_1, J_2, J_3)$ enforces physicality through the Collisionless Boltzmann equation (weak Jeans theorem) and assumes that the samples are mostly regular (\ie non chaotic) with non-resonant frequencies (strong Jeans theorem) \citep{BinneyTremaine}, which is a reasonable assumption for the Milky Way \citep{Michtchenko2017}.

\item  A differentiable density estimator can then be employed to deduce the density function of orbits in actions $f(\vec{J})$. For this purpose, a normalizing flow (\cite{NF_review}) or a diffusion model adapted to tabular data such as \texttt{TabDDPM} (\cite{TabDDPM}) can be used. 

\item  Sampling this function enables us to obtain new orbits $\{\vec{J}_i\}_{i<N_{orbits}}$ in realistic proportions. 

\item  These can in turn be sampled to deduce stellar coordinates in actions and angles:  $\{(\vec{J}, \vec{\theta})_i\}_{i<N_*}$. 

\item  By applying an inverse differentiable transformation, one can obtain the Cartesian coordinates of this augmented and phased-mixed stellar population: $\{(\vec{x}, \vec{v})_i\}_{i<N_*}$.

\item  These can be used to infer a smooth density function in phase-space $f(\vec{x}, \vec{v})$. 

\item  Finally, this density function can be compared to initial observations, using a negative log-likelihood loss function : $\sum_{i=1} ^{n_*} log(f_{\alpha}((\Vec{x},\Vec{v})_i))$.
\item  Since this final density function depends on the potential neural network's parameters $\alpha$ through all of the steps described above, these can be adjusted to minimize this discrepancy. This process can be repeated iteratively until convergence of $\Phi_\alpha$. 

\end{enumerate}

In essence, in our workflow, we are assuming that the system is quasi-stationary (which is well verified within a $\sim 200$ \Myr\ time-scale for the Milky Way \citealt{Hou2015}) and computing the free form potential that stabilizes the observed stellar distribution.

It is worth noting that fitting the large number of parameters that make up neural networks is made possible through backpropagation, which involves computing gradients for each single mathematical operation performed in the workflow. While this approach is powerful, it presents challenges as it necessitates the tracking of all gradients and the utilization of differentiable operations only. We utilize \texttt{PyTorch} \citep{pytorch} for this purpose.\\

We demonstrate the efficacy of our scheme using a toy system, substituting Gaia data with synthetic data from an isochrone whose potential is given by: $\Phi(r) = -{G.M}/{(b+\sqrt{r^2+b^2})}$, using the  analytical canonical transformation to actions and angles (see \cite{BinneyTremaine}) and a simple Gaussian mixture for the density estimation.\\

In this toy showcase, we are able to recover the isochrone potential within a mean relative error of $~0.1\%$ showing that it is possible to use gradients to backpropagate through all of the steps necessary to recover a gravitational field from observations, including an orbit integration, a density estimation, a change of coordinates to actions/angle and a data augmentation using actions. Moreover, we note that our framework can be extended to also leverage the constraints from the numerous stellar streams recently discovered \citep{MW_streams_atlas, charting_acceleration_field_MW_I, Typhon, Antaeus} by requiring the potential to be such that stars from a single stream fall the same orbit $\vec{J}$.

\section{Distilling a neural network into an analytical function}
\label{sec:SymbolicRegression}

Although the agnostic recovery of a neural network $\Phi_\alpha$ enclosing a potential model for the Milky Way would be of enormous value. We note that such a black box model would contrast with usual empirical laws in that it would be very difficult if not impossible to it connect with theory. Therefore, we suggest the use of symbolic regression which consists in the inference of a free-form symbolic analytical function $f: \mathbb{R}^n \longrightarrow \mathbb{R}$ that fits $y = f(\mathbf{x})$ given $(\mathbf{x}, y)$ data for distilling the potential neural network into an intelligible and interpretable analytical function. Symbolic regression is distinct from  numerical parameter optimization procedures in that it consists in a search in the space of functional forms themselves by optimizing the arrangement of mathematical symbols (e.g. $x$, $+$, $-$, $\times$, $/$, $\sin$, $\exp$, $\log$, ...).\\

\begin{figure*}[h]
\begin{center}
\includegraphics[angle=0, clip, width=1.\hsize]{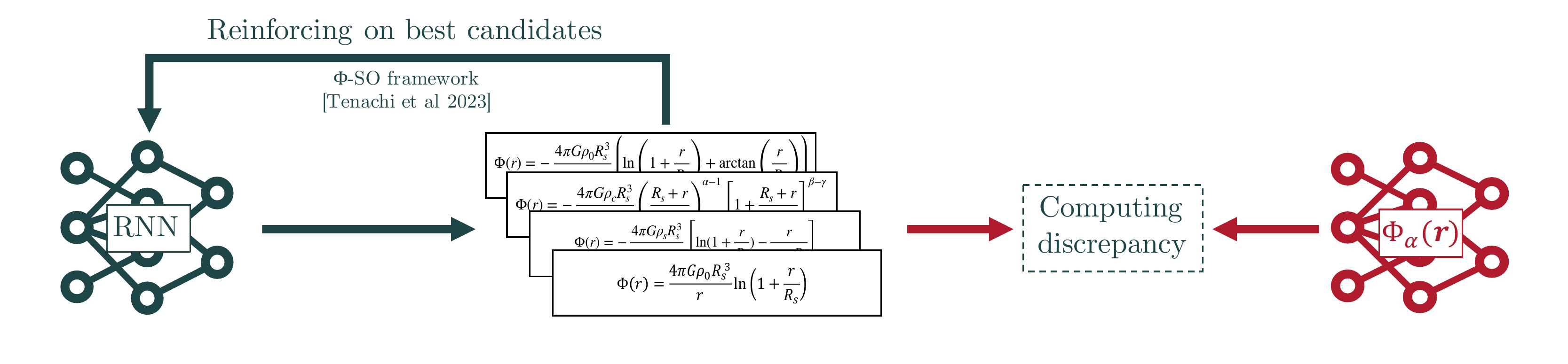}
\end{center}
\caption{Distilling a neural network into a interpretable analytical expression with \PhySO\ \citep{physo}. An RNN generates trial expressions, their ability to reproduce the neural network $\Phi_\alpha$ is assessed and the best ones are reinforced. This process is repeated iteratively until convergence of the RNN and the extraction a set of high quality expressions.}
\label{fig:SR}
\end{figure*}

Here we adopt the Physical Symbolic Optimization (\PhySO) framework detailed in \cite{physo} which was built from the ground up for physics. In this framework, the search space is reduced by leveraging physical units constraints (\eg cos is dimensionless, $\text{velocity} + \text{length}/ \placeholder \iff$ $\placeholder$ is a time etc.) and proposing physically meaningful expressions only. As illustrated in Figure~\ref{fig:SR} \PhySO\ relies on a recurrent neural network (RNN) to generate multiple trial analytical expressions. Fit quality of these expressions can then be assessed against data generated using the $\Phi_\alpha$ neural network. Best expressions are then reinforced and the process is repeated until the RNN converges and a set of high quality expressions that reproduce $\Phi_\alpha$ predictions is obtained. We note that since this framework relies on reinforcement learning, in addition to fit quality any criteria (even non-differentiable ones) can be used including the condition: $\displaystyle\lim_{r\to\infty} \Phi(r) = 0$. Using the \PhySO\ framework, we are able to successfully recover the potential of the toy isochrone system described in Section \ref{sec:MassFinder}.

\section*{Code availability}
\label{sec:availability}
The documented code for the \PhySO\ algorithm along with demonstration notebooks are available on GitHub \href{https://github.com/WassimTenachi/PhySO}{github.com/WassimTenachi/PhySO} \github{https://github.com/WassimTenachi/PhySO}.

\section*{Acknowledgments}
RI acknowledges funding from the European Research Council (ERC) under the European Unions Horizon 2020 research and innovation programme (grant agreement No. 834148).

\bibliography{MLPotential}

\begin{thebibliography}{}
\expandafter\ifx\csname natexlab\endcsname\relax\def\natexlab#1{#1}\fi
\providecommand{\url}[1]{\href{#1}{#1}}
\providecommand{\dodoi}[1]{doi:~\href{http://doi.org/#1}{\nolinkurl{#1}}}
\providecommand{\doeprint}[1]{\href{http://ascl.net/#1}{\nolinkurl{http://ascl.net/#1}}}
\providecommand{\doarXiv}[1]{\href{https://arxiv.org/abs/#1}{\nolinkurl{https://arxiv.org/abs/#1}}}

\bibitem[{Binney \& Tremaine(2011)}]{BinneyTremaine}
Binney, J., \& Tremaine, S. 2011, Galactic dynamics, Vol.~13 (Princeton
  university press)

\bibitem[{Bullock \& Boylan-Kolchin(2017)}]{Bullock_LCDM_challenges}
Bullock, J.~S., \& Boylan-Kolchin, M. 2017, ARAA, 55, 343,
  \dodoi{10.1146/annurev-astro-091916-055313}

\bibitem[{{Gaia Collaboration}(2022)}]{GaiaDR3}
{Gaia Collaboration}. 2022, A\&A, \dodoi{10.1051/0004-6361/202243940}

\bibitem[{Hou \& Han(2015)}]{Hou2015}
Hou, L.~G., \& Han, J.~L. 2015, MNRAS, 454, 626, \dodoi{10.1093/mnras/stv1904}

\bibitem[{Ibata {et~al.}(2021{\natexlab{a}})Ibata, Diakogiannis, Famaey, \&
  Monari}]{ActionFinder}
Ibata, R., Diakogiannis, F.~I., Famaey, B., \& Monari, G. 2021{\natexlab{a}},
  ApJ, 915, 5, \dodoi{10.3847/1538-4357/abfda9}

\bibitem[{Ibata {et~al.}(2021{\natexlab{b}})Ibata, Malhan, Martin, Aubert,
  Famaey, Bianchini, Monari, Siebert, Thomas, Bellazzini, Bonifacio, Caffau, \&
  Renaud}]{charting_acceleration_field_MW_I}
Ibata, R., Malhan, K., Martin, N., {et~al.} 2021{\natexlab{b}}, ApJ, 914, 123,
  \dodoi{10.3847/1538-4357/abfcc2}

\bibitem[{{Kotelnikov} {et~al.}(2022){Kotelnikov}, {Baranchuk}, {Rubachev}, \&
  {Babenko}}]{TabDDPM}
{Kotelnikov}, A., {Baranchuk}, D., {Rubachev}, I., \& {Babenko}, A. 2022, arXiv
  e-prints, arXiv:2209.15421, \dodoi{10.48550/arXiv.2209.15421}

\bibitem[{Malhan {et~al.}(2022)Malhan, Ibata, Sharma, Famaey, Bellazzini,
  Carlberg, D’Souza, Yuan, Martin, \& Thomas}]{MW_streams_atlas}
Malhan, K., Ibata, R.~A., Sharma, S., {et~al.} 2022, ApJ, 926, 107,
  \dodoi{10.3847/1538-4357/ac4d2a}

\bibitem[{{Michtchenko} {et~al.}(2017){Michtchenko}, {Vieira}, {Barros}, \&
  {L{\'e}pine}}]{Michtchenko2017}
{Michtchenko}, T.~A., {Vieira}, R.~S.~S., {Barros}, D.~A., \& {L{\'e}pine},
  J.~R.~D. 2017, A\&A, 597, A39, \dodoi{10.1051/0004-6361/201628895}

\bibitem[{Oria {et~al.}(2022)Oria, Tenachi, Ibata, Famaey, Yuan, Arentsen,
  Martin, \& Viswanathan}]{Antaeus}
Oria, P.-A., Tenachi, W., Ibata, R., {et~al.} 2022, ApJL, 936, L3,
  \dodoi{10.3847/2041-8213/ac86d3}

\bibitem[{Papamakarios {et~al.}(2021)Papamakarios, Nalisnick, Rezende, Mohamed,
  \& Lakshminarayanan}]{NF_review}
Papamakarios, G., Nalisnick, E., Rezende, D.~J., Mohamed, S., \&
  Lakshminarayanan, B. 2021, JMLR, 22, \dodoi{10.48550/arXiv.1912.02762}

\bibitem[{Paszke {et~al.}(2019)Paszke, Gross, Massa, Lerer, Bradbury, Chanan,
  Killeen, Lin, Gimelshein, Antiga, {et~al.}}]{pytorch}
Paszke, A., Gross, S., Massa, F., {et~al.} 2019, NeurIPS, 32,
  \dodoi{10.48550/arXiv.1912.01703}

\bibitem[{{Tenachi} {et~al.}(2023){Tenachi}, {Ibata}, \&
  {Diakogiannis}}]{physo}
{Tenachi}, W., {Ibata}, R., \& {Diakogiannis}, F.~I. 2023, arXiv e-prints,
  arXiv:2303.03192, \dodoi{10.48550/arXiv.2303.03192}

\bibitem[{Tenachi {et~al.}(2022)Tenachi, Oria, Ibata, Famaey, Yuan, Arentsen,
  Martin, \& Viswanathan}]{Typhon}
Tenachi, W., Oria, P.-A., Ibata, R., {et~al.} 2022, ApJL, 935, L22,
  \dodoi{10.3847/2041-8213/ac874f}

\end{thebibliography}
\bibliographystyle{aasjournal}

\end{document}